\documentclass[12pt]{article}

\usepackage{amsmath}
\usepackage{amsfonts}
\usepackage{amssymb}

\usepackage{mathstyle}
\usepackage{flexisym}
\usepackage{breqn}

\usepackage{srcltx}

\usepackage{pgf}
\input xy
\xyoption{all}

\usepackage{hyperref}

\newtheorem{theo}{Theorem }

\newtheorem{rema}{Remark}

\def\F{\mathcal{F}}

\def\R{\mathbb{R}}

\def\P{\mathbb{P}}

\def\al{\alpha}

\def\la{\lambda}

\title{Rational interpolants and solutions of dispersionless Hirota system}
\author{
Andriy Panasyuk\\
Faculty of Mathematical and Natural Sciences\\
Cardinal Stefan Wyszyński University\\
Warsaw,  Poland\\
}
\date{}
\begin{document}
\bibliographystyle{plain}

\maketitle

\begin{abstract}
The aim of this paper is to construct a class of explicit nontrivial rational solutions of the dispersionless Hirota system of PDEs.  All the solutions in this class are of homogeneity degree 1 and are quotients of homogeneous polynomials. It is well-known that the solutions of the Hirota dispersionless systems describe Veronese webs. By nontriviality of the solutions it is meant that the corresponding  Veronese webs  are nonflat at generic points.
\end{abstract}

\section{Introduction}
A  one-parametric family of foliations $\{\F_\la\}$ of codimension one in a $n$-dimensional space is called a \emph{Veronese web} \cite{gz2} if in a vicinity of any point there exist a local coframe $\al_0,\ldots,\al_{n-1}$ such that the corresponding annihilating 1-form $\al^\la$, $T\F_\la=\ker\al^\la$, is a polynomial $\al_0+\la\al_1  +\cdots+\la^{n-1}\al_{n-1}$ of order $n-1$ in $\la$. A Veronese web is \emph{flat} if in a vicinity of any point one can find a local system of coordinates $x_i$ such that $T\F_\la=\ker(dx_0+\la dx_1+\cdots+\la^{n-1} dx_{n-1})$, i.e. if locally the foliations $\F_\la$ are simultaneously equivalent to the foliations of parallel hypersurfaces. Veronese webs were introduced in the paper cited as a tool for the local study of the  so-called bihamiltonian systems of ODEs. It turns out that there exist nonflat Veronese webs and their description is an important geometric and analytic problem.

In a seminal paper \cite{z2} I. Zakharevich studied a nonlinear PDE of the form
\begin{equation}\label{e2}
(\la_2-\la_3)f_1 f_{23}+(\la_3-\la_1)f_2f_{31}+(\la_1-\la_2)f_3f_{12}=0
\end{equation}
which nowadays is commonly known as \emph{dispersionless Hirota equation} (here $f_{i}:=\frac{\partial f}{\partial x_i}$ and $f_{ij}:=\frac{\partial^2 f}{\partial x_i\partial x_j}$ and $\la_i$, $i=1,2,3$, are arbitrary pairwise distinct parameters). Its solutions describe   Veronese webs in 3D. More precisely, equation (\ref{e2}) is equivalent to the Frobenius integrability condition
\begin{equation}\label{frob}
d\al^\la\wedge\al^\la\equiv_\la 0
\end{equation}
 for the one-form
\begin{equation}\label{vero}
\al^\la=(\la-\la_1)(\la-\la_2)(\la-\la_3)\left(\frac{f_1dx_1}{\la-\la_1} +\frac{f_2dx_2}{\la-\la_2}+
\frac{f_3dx_3}{\la-\la_3}\right)
\end{equation}
which annihilates the corresponing foliation $\F_\la$. This construction can be easily generalized to higher dimensions. The corresponding system of PDEs, which will be \emph{called dispersionless Hirota system},  is equivalent to condition (\ref{frob}) for the one-form
\begin{equation}\label{vero2}
\al^\la=\prod_{i=1}^{n}(\la-\la_i)\left(\sum_{i=1}^{n}\frac{f_idx_i}{\la-\la_i} \right),
\end{equation}
where now $(x_1,\ldots, x_n)$ are coordinates in a $n$-dimensional space and $\la_i$, $i=1,\ldots,n$, are arbitrary pairwise distinct parameters.
Explicitly this system looks as
\begin{equation}\label{hiro}
(\la_j-\la_k)f_i f_{jk}+(\la_k-\la_i)f_jf_{ki}+(\la_i-\la_j)f_kf_{ij}=0,
\end{equation}
where the indices $i,j,k$ exhaust all the triples of pairwise distinct elements from the set $\{1,\ldots,n\}$\footnote{Note that starting from dimension 4 the full set of equations is algebraically dependent. For instance in 4D any of the four equations is an algebraic consequence of the remaining three ones.}.
The one-form (\ref{vero2}) annihilates a distribution of codimension one for any $\la$ and the solutions of  system (\ref{hiro}) describe $n$-dimensional Veronese webs.

The aim of this short note is to construct a class of explicit rational solutions of system (\ref{hiro}). Recall \cite{Cauchy,cuytWuytack} that a \emph{rational interpolant} or \emph{Cauchy interpolant} of \emph{order} $[k/l]$, $k+l+1=n$, with\emph{ nodes} $\la_1,\ldots,\la_n$, $\la_i\not=\la_j$, $i\not=j$, and \emph{values} $x_i$, $x_i\in\R$, is a rational function
\begin{equation}\label{interp}
F(\la):=\frac{p(\la)}{q(\la)}:=\frac{p_0+p_1\la+\cdots+p_k\la^k}{1+q_1\la+\cdots+q_l\la^l}
\end{equation}
such that $F(\la_i)=x_i$ for any $i=1,\ldots,n$. The system $F(\la_i)=x_i$ is a linear system of $n$ equations on $n$ unknowns $p_0,\ldots,q_l$ and has a unique solution. It is given by $p(\la)=P(\la)/Q(0)$ and $q(\la)=Q(\la)/Q(0)$ \cite{Jacobi}, \cite[Prop. 2.1]{doliwa}, where
\begin{equation*}
  P(\la)=\left|\begin{array}{cccccccc}
               1 & \la_1 & \ldots & \la_1^k & -x_1 & -x_1\la_1 & \ldots & -x_1\la_1^l \\
               \vdots & \vdots &  & \vdots & \vdots & \vdots &   & \vdots \\
               1 & \la_n & \ldots & \la_n^k & -x_n & -x_n\la_n & \ldots & -x_n\la_n^l\\
               1 & \la & \cdots & \la^k & 0 & 0 & \cdots & 0
             \end{array}\right|
\end{equation*}
and
\begin{equation*}
  Q(\la)=\left|\begin{array}{cccccccc}
               1 & \la_1 & \ldots & \la_1^k & -x_1 & -x_1\la_1 & \ldots & -x_1\la_1^l \\
               \vdots & \vdots &  & \vdots & \vdots & \vdots &   & \vdots \\
               1 & \la_n & \ldots & \la_n^k & -x_n & -x_n\la_n & \ldots & -x_n\la_n^l\\
               0 & 0 & \cdots & 0 & 1 & \la & \cdots & \la^l
             \end{array}\right|.
\end{equation*}
In particular, put
\begin{equation}\label{pp}
P_k=(-1)^{n+k}\left|\begin{array}{cccccccc}
               1 & \la_1 & \ldots & \la_1^{k-1} & -x_1 & -x_1\la_1 & \ldots & -x_1\la_1^l \\
               \vdots & \vdots &  & \vdots & \vdots & \vdots &   & \vdots \\
               1 & \la_n & \ldots & \la_n^{k-1} & -x_n & -x_n\la_n & \ldots & -x_n\la_n^l\\
                            \end{array}\right|
\end{equation}
and
\begin{equation}\label{qq}
Q_l=(-1)^{n+k+l+1}\left|\begin{array}{cccccccc}
               1 & \la_1 & \ldots & \la_1^k & -x_1 & -x_1\la_1 & \ldots & -x_1\la_1^{l-1} \\
               \vdots & \vdots &  & \vdots & \vdots & \vdots &   & \vdots \\
               1 & \la_n & \ldots & \la_n^k & -x_n & -x_n\la_n & \ldots & -x_n\la_n^{l-1}\\
                           \end{array}\right|.
\end{equation}
for the highest coefficients.
Below we shall prove the following theorem.
\begin{theo}\label{th}
Let $F(\la,x)=\frac{p_0(x)+p_1(x)\la+\cdots+p_k(x)\la^k}{1+q_1(x)\la+\cdots+q_l(x)\la^l}$, $x:=(x_1,\ldots,x_n)$, be the Cauchy interpolant with { nodes} $\la_1,\ldots,\la_n$, $\la_i\not=\la_j$, $i\not=j$, and  {values} $x_i$.  Then the function $f(x):=\frac{p_k(x)}{q_l(x)}=\frac{P_k(x)}{Q_l(x)}$ is a solution to system (\ref{hiro}). If $k>0$ and $l>0$, the  corresponding Veronese web is nonflat at generic point.
\end{theo}

\begin{rema}\label{rem}\rm
In the case $l=0$ the Cauchy interpolation problem degenerates to the Lagrange interpolation problem. The corresponding Veronese web is flat in this case (the coefficients $p_0(x),\ldots,p_k(x)$ play the role of coordinates used in the definition of flatness).
\end{rema}
\begin{rema}\label{rem2}\rm
Also in the case $k=0$ the corresponding Veronese web is flat. Indeed, the corresponding Cauchy interpolation problem is equivalent to the Lagrange interpolation problem for the polynomial $\frac1{p_0}+\frac{q_1}{p_0}\la+\cdots+\frac{q_l}{p_0}\la^l$ and nodes $\frac1{x_i}$.
 \end{rema}
 \begin{rema}\label{rem3}\rm
 It is easy to see that, if $f(x_1,\ldots,x_n)$ is a solution to the Hirota system (\ref{hiro}), then so is any function $\Phi(f(\varphi_1(x_1),\ldots,\varphi_n(x_n))$ with smooth $\Phi(t),\varphi_i(t)$. The cases of orders $[k/l]$ and $[l/k]$ with $k,l>0$ are related by $\Phi(t)=1/t,\varphi_i(t)=1/t$.
\end{rema}

\section{Veronese curves, Veronese webs, and the proof of the theorem}

Recall that a \emph{Veronese curve} (or a \emph{rational normal curve}) is a map $c:\P^1(\R)\to \P^{n-1}(\R)$ that in some homogeneous coordinates of projective space can be given by $[\mu:\la]\mapsto [\mu^{n-1}\la^0:\mu^{n-2}\la^1:\cdots:\mu^0\la^{n-1}]$, or $\la\mapsto(1,\la,\ldots,\la^{n-1})$ in the corresponding affine chart of $\P^1$ and underlying linear space $\R^{n}$ of $\P^{n-1}$. The crucial is the following uniqueness property of the Veronese curve: for any pairwise distinct $\la_0,\ldots\la_n\in\P^1$ and any $v_0,\ldots,v_n\in\P^n$ in general position there exists a unique Veronese curve $c$ such that $c(\la_i)=v_i$, $i=0,\ldots,n$. If $\la_0=\infty$ and $V_i\in\R^{n}$,  are any vectors such that $v_i=p(V_i)$, $i=1,\ldots,n$, where $p:\R^{n}\to \P^{n-1}$ is the canonical projection, then the formula
$$
c(\la)=p\left(\prod_{i=1}^{n}(\la-\la_i)\sum_{i=1}^{n}\frac{V_i}{\la-\la_i}\right).
$$
gives the unique Veronese curve $c$ such that $c(\la_i)=v_i=p(V_i)$, $i=1,\ldots,n$, and $c(\la_0)=c(\infty)=p(V_1+\cdots+V_n)$.

In particular, assuming that $f_i(x)\not=0$ for a fixed $x\in \R^n$, one-form (\ref{vero2}) represents the unique Veronese curve $c$ in $\P(T_x^*\R^n)$ such that $c(\la_i)=p(dx_i)$, $i=1,\ldots,n$, and $c(\infty)=p(df)$. If moreover one-form (\ref{vero2}) is Frobenius integrable, then its kernel represents a Veronese web $\{\F_\la\}$ with the following property:
\begin{equation}\label{prop}
\F_{\la_i}=\{x_i=const\}, i=1,\ldots,n,\F_\infty=\{f=const\}.
\end{equation}

From this it follows that, if one is able to construct a Veronese web $\{\F_\la\}$ with property (\ref{prop}) with some smooth function $f$, then this function will satisfy system (\ref{hiro}). Indeed, the one-form $\al^\la$ given by (\ref{vero2}) by uniqueness will satisfy $\ker\al^\la=T\F_\la$ for any $\lambda$ and, since $T\F_\la$ is an integrable distribution, $\al^\la$ will be Frobenius integrable and (\ref{hiro}) is satisfied by $f$.

Now we shall construct a Veronese web with property (\ref{prop}) with $f(x)=\frac{p_k(x)}{q_l(x)}$. Let $\F_\la=\{F(x,\la)=const\}$ be the foliation cut by the Cauchy interpolant. Then, since the one-form (we skip the arguments of the functions $p_i(x),q_j(x)$ for brevity)
\begin{eqnarray*}
dF(\la,x)=\frac1{(1+q_1\la+\cdots+q_l\la^l)^2}((1+q_1\la+\cdots+q_l\la^l)
d(p_0+p_1\la+\cdots+p_k\la^k)-\\
(p_0+p_1\la+\cdots+p_k\la^k)d(1+q_1\la+\cdots+q_l\la^l))
\end{eqnarray*}
up to a nonzero factor is a polynomial of order $k+l=n-1$, the family $\{\F_\la\}$ is a Veronese web. The fact that $\F_\infty=\{f=const\}$ implies  that $f$ satisfies (\ref{hiro}).

To finish the proof of the theorem we have to show that the corresponding Veronese web is nonflat. To this end we shall use the following criterion \cite[Ch. I(II), Prop. 6]{rigal}: a Veronese web $\{\F_\la\}$ given by the annihilating form $\al^\la=\al_0+\la\al_1  +\cdots+\la^{n-1}\al_{n-1}$ is flat if and only if the one-form $\al_1$ or $\al_{n-2}$ is Frobenius integrable.

In our case we have $\al_1=dp_1+q_1dp_0-p_0dq_1$ and $d\al_1\wedge\al_1=2dq_1\wedge dp_0\wedge dp_1$.
The functional correspondence $(p_0,\ldots,p_k,q_1,\ldots,q_l)\leftrightarrow (x_1,\ldots,x_n)$ is invertible at generic point, hence the functions $p_0,\ldots,p_k,q_1,\ldots,q_l$ are functionally independent at generic point and $d\al_1\wedge\al_1\not=0$. This finishes the proof.

\section{Examples}

It is enough to consider only cases $k\ge l$ (cf. Remark \ref{rem3}).

In dimension 3 we have the only possibility leading to a nonflat case: $k=l=1$. Explicitly,
$$
f(x)=\frac{p_1(x)}{q_1(x)}=\frac{(\la_1-\la_2)x_1 x_2+(\la_2-\la_3)x_2 x_3+(\la_3-\la_1)x_3 x_1}{(\la_3-\la_2)x_1+(\la_1-\la_3) x_2+(\la_2-\la_1)x_3}.
$$
In dimension 4 the case $k=2,l=1$ gives
\begin{eqnarray}\label{4D}
\nonumber  p_2(x)=(\la_3^2-\la_4^2)(\la_1-\la_2)x_1 x_2-(\la_2^2-\la_4^2)(\la_1-\la_3)x_1 x_3+(\la_2^2-\la_3^2)(\la_1-\la_4)x_1 x_4+\\
\nonumber (\la_1^2-\la_4^2)(\la_2-\la_3)x_2 x_3-(\la_1^2-\la_3^2)(\la_2-\la_4)x_2 x_4+(\la_1^2-\la_2^2)(\la_3-\la_4)x_3 x_4,\\
\nonumber q_1(x)=(\la_3-\la_4)(\la_2-\la_4)(\la_2-\la_3) x_1-(\la_3-\la_4)(\la_2-\la_4)(\la_2-\la_3) x_2+\\
(\la_2-\la_4)(\la_1-\la_4)(\la_1-\la_2) x_3-(\la_2-\la_3)(\la_1-\la_3)(\la_1-\la_2) x_4.
\end{eqnarray}
In dimension 5 we have two nontrivial possibilities. For the case $k=3,l=1$ we have
\begin{eqnarray*}
p_3(x)=(\la_4-\la_5)(\la_3-\la_5)(\la_3-\la_4)(\la_1-\la_2)x_1 x_2
-(\la_4-\la_5)(\la_2-\la_5)(\la_2-\la_4)(\la_1-\la_3)x_1 x_3\\
+(\la_3-\la_5)(\la_2-\la_5)(\la_2-\la_3)(\la_1-\la_4)x_1 x_4
-(\la_3-\la_4)(\la_2-\la_4)(\la_2-\la_3)(\la_1-\la_5)x_1 x_5\\
+(\la_4-\la_5)(\la_1-\la_5)(\la_1-\la_4)(\la_2-\la_3)x_2 x_3
-(\la_3-\la_5)(\la_1-\la_5)(\la_1-\la_3)(\la_2-\la_4)x_2 x_4\\
+(\la_3-\la_4)(\la_1-\la_4)(\la_1-\la_3)(\la_2-\la_5)x_2 x_5
+(\la_2-\la_5)(\la_1-\la_5)(\la_1-\la_2)(\la_3-\la_4)x_3 x_4\\
-(\la_2-\la_4)(\la_1-\la_4)(\la_1-\la_2)(\la_3-\la_5)x_3 x_5
+(\la_2-\la_3)(\la_1-\la_3)(\la_1-\la_2)(\la_4-\la_5)x_4 x_5,\\
q_1(x)=-(\la_4-\la_5)(\la_3-\la_5)(\la_3-\la_4)(\la_2-\la_5)(\la_2-\la_4)(\la_2-\la_3) x_1\\
+(\la_4-\la_5)(\la_3-\la_5)(\la_3-\la_4)(\la_1-\la_5)(\la_1-\la_4)(\la_1-\la_3) x_2\\
-(\la_4-\la_5)(\la_2-\la_5)(\la_2-\la_4)(\la_1-\la_5)(\la_1-\la_4)(\la_1-\la_2) x_3\\
+(\la_3-\la_5)(\la_2-\la_5)(\la_2-\la_3)(\la_1-\la_5)(\la_1-\la_3)(\la_1-\la_2) x_4\\
-(\la_3-\la_4)(\la_2-\la_4)(\la_2-\la_3)(\la_1-\la_4)(\la_1-\la_3)(\la_1-\la_2) x_5.
\end{eqnarray*}
For the second possibility, $k=l=2$, we get
\begin{eqnarray*}
p_2(x)=-(\la_4-\la_5)(\la_2-\la_3)(\la_1-\la_3)(\la_1-\la_2)x_1 x_2 x_3
+(\la_3-\la_5)(\la_2-\la_4)(\la_1-\la_4)(\la_1-\la_2)x_1 x_2 x_4\\
-(\la_3-\la_4)(\la_2-\la_5)(\la_1-\la_5)(\la_1-\la_2)x_1 x_2 x_5
-(\la_2-\la_5)(\la_3-\la_4)(\la_1-\la_4)(\la_1-\la_3)x_1 x_3 x_4\\
+(\la_2-\la_4)(\la_3-\la_5)(\la_1-\la_5)(\la_1-\la_3)x_1 x_3 x_5
-(\la_2-\la_3)(\la_4-\la_5)(\la_1-\la_5)(\la_1-\la_4)x_1 x_4 x_5\\
+(\la_1-\la_5)(\la_3-\la_4)(\la_2-\la_4)(\la_2-\la_3)x_2 x_3 x_4
-(\la_1-\la_4)(\la_3-\la_5)(\la_2-\la_5)(\la_2-\la_3)x_2 x_3 x_5\\
+(\la_1-\la_3)(\la_4-\la_5)(\la_2-\la_5)(\la_2-\la_4)x_2 x_4 x_5
-(\la_1-\la_2)(\la_4-\la_5)(\la_3-\la_5)(\la_3-\la_4)x_3 x_4 x_5,\\
q_2(x)=-(\la_4-\la_5)(\la_3-\la_5)(\la_3-\la_4)(\la_1-\la_2) x_1 x_2+(\la_4-\la_5)(\la_2-\la_5)(\la_2-\la_4)(\la_1-\la_3) x_1 x_3\\
-(\la_3-\la_5)(\la_2-\la_5)(\la_2-\la_3)(\la_1-\la_4) x_1 x_4+(\la_3-\la_4)(\la_2-\la_4)(\la_2-\la_3)(\la_1-\la_5) x_1 x_5\\
-(\la_4-\la_5)(\la_1-\la_5)(\la_1-\la_4)(\la_2-\la_3) x_2 x_3
+(\la_3-\la_5)(\la_1-\la_5)(\la_1-\la_3)(\la_2-\la_4) x_2 x_4\\
-(\la_3-\la_4)(\la_1-\la_4)(\la_1-\la_3)(\la_2-\la_5) x_2 x_5
-(\la_2-\la_5)(\la_1-\la_5)(\la_1-\la_2)(\la_3-\la_4) x_3 x_4\\
+(\la_2-\la_4)(\la_1-\la_4)(\la_1-\la_2)(\la_3-\la_5) x_3 x_5
-(\la_2-\la_3)(\la_1-\la_3)(\la_1-\la_2)(\la_4-\la_5) x_4 x_5.
\end{eqnarray*}

It follows from formulas (\ref{pp}) and (\ref{qq}) that all the solutions of the Hirota system of the form $p_k(x)/q_l(x)$, in particular those above, have the following properties:
\begin{enumerate}\item $p_k(x)$ and $q_l(x)$ are homogeneous polynomials in $x$;
\item $\deg(p_k(x))=\deg(q_l(x))+1$;
\item the coefficients of the polynomials $p_k(x)$ and $q_l(x)$ sum up to zero (this can be seen by substituting $x=(1,\ldots,1)$ to (\ref{pp}) and (\ref{qq})).\end{enumerate}
We conclude this section by mentioning that from the solutions $p_k(x)/q_l(x)$ one can also construct rational solutions satisfying condition 2 but not satisfying 1 or 3. This observation is based on the known fact that the restriction of a Veronese web to any of its leaves is again a Veronese web. In our construction all the coordinate hypersurfaces $\{x_i=const\}$ are the leaves of the Veronese web given by the annihilating form (\ref{vero2}). Another emanation of this fact is that any solution of system (\ref{hiro}) after the restriction to the hypersurface $\{x_i=const\}$ is a solution to the lower dimensional sytem in which the corresponding coordinate is not present (this can be seen also directly from the form of (\ref{hiro})). This process of restriction can be performed repeatedly.

For instance, putting $x_4=a=const$ in formulas (\ref{4D}) one gets a solution of equation (\ref{e2}). If $a=0$ conditions 1 and 2 are satisfied but condition 3 is violated. If $a\not=0$ the homogeneity is also lost.

\section{Concluding remarks}

In \cite{pHirota} several classes of PDEs  are considered, which also describe Veronese webs but are contactly inequivalent to (\ref{e2}). W. Krynski \cite{KrynskiDef} interpreted some of these PDEs from twistorial point of view as deformations of equation (\ref{e2}) which informally can be understood as the limiting cases when two of the  parameters $\la_1,\la_2,\la_3$ or all of them tend to one point. Analogous deformations and their interpretation are possible also for higher-dimensional cases of system (\ref{hiro}).

We conclude this paper by remarking that a class of solutions similar to that studied above should exist also for the deformed Hirota systems. The Cauchy interpolation problem  should be replaced by the Pad\'{e} approximation problem, i.e. by seeking for rational functions $F(\la)$ of the form (\ref{interp}) such that for some fixed $\la_0$ one has $\frac{d^i}{d\la^i}|_{\la=\la_0}F(\la)=x_i$, $i=0,\ldots,n-1$  (if one aims in the solution of the PDE corresponding to ``gluing'' all the parameters $\la_i$), or by mixed interpolation-approximation problem (for ''partial gluing'').

\end{document}